\documentclass[aps,prl,superscriptaddress,preprintnumbers,twocolumn,amsmath,amssymb,floatfix]{revtex4-1}
\usepackage{graphicx}
\usepackage{color}
\definecolor{RED}{rgb}{1,0,0}
\definecolor{BLUE}{rgb}{0,0,1}
\begin{document}

\title{Band dependent inter-layer $f$-electron hybridization in CeRhIn$_5$}
\author{Q. Y. Chen}
\affiliation{State Key Laboratory of Surface Physics and Department of Physics, Fudan University, Shanghai 200433, China}
\affiliation{Science and Technology on Surface Physics and Chemistry Laboratory, Mianyang 621908, China}
\author{D. F. Xu}
\author{X. H. Niu}
\author{R. Peng}
\author{H. C. Xu}
\author{C. H. P. Wen}
\author{X. Liu}
\author{L. Shu}
\affiliation{State Key Laboratory of Surface Physics and Department of Physics, Fudan University, Shanghai 200433, China}
\author{S. Y. Tan}
\author{X. C. Lai}
\affiliation{Science and Technology on Surface Physics and Chemistry Laboratory, Mianyang 621908, China}
\author{Y. J. Zhang}
\affiliation{Center for Correlated Matter, Zhejiang University, Hangzhou, 310058, China}
\affiliation{Department of Physics, Zhejiang University, Hangzhou, 310027, China}
\author{H. Lee}
\affiliation{Center for Correlated Matter, Zhejiang University, Hangzhou, 310058, China}
\author{V. N. Strocov}
\author{F. Bisti}
\affiliation{Swiss Light Source, Paul Scherrer Institute, CH-5232 Villigen, Switzerland}
\author{P. Dudin}
\affiliation{Diamond Light Source, Harwell Science and Innovation Campus, Didcot OX11 0DE, United Kingdom}
\author{J.-X. Zhu}
\affiliation{Theoretical Division and Center for Integrated Nanotechnologies, Los Alamos National Laboratory, Los Alamos, NM 87545}
\author{H. Q. Yuan}
\affiliation{Center for Correlated Matter, Zhejiang University, Hangzhou, 310058, China}
\affiliation{Department of Physics, Zhejiang University, Hangzhou, 310027, China}
\affiliation{Collaborative Innovation Center of Advanced Microstructures, Nanjing 210093, China}
\author{S. Kirchner}\email{stefan.kirchner@correlated-matter.com}
\affiliation{Center for Correlated Matter, Zhejiang University, Hangzhou, 310058, China}
\author{D. L. Feng}
\email{dlfeng@fudan.edu.cn}
\affiliation{State Key Laboratory of Surface Physics and Department of Physics, Fudan University, Shanghai 200433, China}
\affiliation{Collaborative Innovation Center of Advanced Microstructures, Nanjing 210093, China}

\begin{abstract}
A key issue in heavy fermion research is how subtle changes in the hybridization between the 4$f$ (5$f$) and conduction electrons can result in fundamentally  different  ground states.
CeRhIn$_5$ stands out as  a particularly notable example:  replacing Rh by either Co or Ir,
antiferromagnetism gives way to superconductivity.
In this photoemission study of CeRhIn$_5$, we demonstrate that the use of resonant ARPES with polarized light allows to extract detailed information on the 4$f$ crystal field states and details on the 4$f$ and conduction electron  hybridization which together determine the ground state.  We
directly observe weakly dispersive Kondo resonances of $f$-electrons and identify two of the three Ce $4f_{5/2}^{1}$ crystal-electric-field levels and band-dependent hybridization, which signals that the hybridization occurs primarily between the Ce $4f$ states in the CeIn$_3$ layer and two more three-dimensional bands composed of the  Rh $4d$ and In $5p$ orbitals in the RhIn$_2$ layer. Our results allow to connect the properties observed at elevated temperatures with the unusual low-temperature properties of this enigmatic heavy fermion compound.
\end{abstract}

\maketitle
It is well known that  the coupling of $f$ electrons in intermetallic rare earth (and actinide) compounds  to  conduction electrons can give rise to Kondo scattering processes, which commonly results in narrow bands near the Fermi energy ($E_F$). The correspondingly large effective mass and  heat capacity over temperature ($T$) ratio at low $T$ gave this materials class its name: heavy fermions \cite{Stewart.84}.
The great variety of possible ground states occurring in these materials  has made them an ideal testbed for exploring strong correlations. As a result, a global phase diagram of heavy fermion systems has started to emerge~\cite{Coleman.10,Si.14} that is considerably richer  than originally anticipated~\cite{Doniach.77}.

The Ce$M$In$_5$ heavy electron family,  whose members are formed out  of alternating $M$In$_2$ ($M$ = Co, Rh, Ir) and CeIn$_3$ layers, typifies this richness \cite{Sidorov.02,Mito.03}. Both CeCoIn$_5$ and CeIrIn$_5$ are superconductors, with $T_c$'s of 2.3~K and 0.4~K, respectively \cite{Petrovic.01,Petrovic.01b}, while CeRhIn$_5$ is an antiferromagnet with predominantly localized moments and a N\'{e}el temperature $T_N$=3.8 K at ambient pressure \cite{Hegger.00}. Rh lies between Co and Ir in the Group VIII elements suggesting that already minute changes can yield  drastically different ground states.  This does raise the question  what
determines the location of a compound in the global phase diagram and thus the ground state?
For CeCoIn$_5$  the quasiparticle bands have been directly observed by angle-resolved photoemission spectroscopy (ARPES), which indicates that the $4f$ electrons are delocalized and participate in the electronic properties at sufficiently low $T$ \cite{Fujimori.06,Koitzsch.13,Koitzsch.08,Qiuyun.17}.
Dynamical mean-field theory (DMFT) calculations for the quasiparticle spectral weight suggest that  while the Ce 4$f$ electrons in CeRhIn$_5$ are more localized than in CeCoIn$_5$, they  still undergo Kondo screening with a subsequent Fermi volume enlargement at low $T$ \cite{Haule.10} suggesting that the ground state of both compounds possesses a  large Fermi surface (FS) which includes the Ce 4$f$ electrons.
In contrast,  de~Haas-van~Alphen (dHvA) oscillations for CeRhIn$_5$ are essentially the same as those for LaRhIn$_5$, indicating that the Ce 4$f$ electrons are localized \cite{Shishido.02,Harrison.04}, which is supported by optical conductivity data \cite{Mena.05}. This is  interpreted as an absence of Kondo screening in CeRhIn$_5$ at ambient pressure and is consistent with the occurrence of unconventional quantum criticality featuring critical Kondo destruction \cite{Shishido2005,Coleman.01,Si2001}.
This interpretation has, however, been challenged in Ref. \cite{Miyake.06}.
In addition,  non-resonant ARPES investigations of  CeRhIn$_5$ find that 4$f$ electrons are predominantly itinerant \cite{Moorea.02}, while another ARPES study suggests that they are \textit{nearly} localized  in the paramagnetic state \cite{Fujimori.03} and a recent scanning tunneling spectroscopy (STS) study reported absence of a hybridization gap for CeRhIn$_5$~\cite{Aynajian.12}.
How can one reconcile these findings? Differences between members of the 115 family can arise from variations in the conduction (or $c$-) electron band structure and  the strength of the 4$f$ multiplet-$c$-electron hybridization. This determines the crystal electric field (CEF) splittings and the Kondo scale $T_K$, which affect the low-energy properties of dense Kondo systems~\cite{Qiuyun.17}.

Here we focus on the temperature range between 180 and 8 K above the N\'eel temperature. Combining  bulk-sensitive soft x-ray ARPES, resonant ARPES, and density functional theory (DFT) calculations, we
i) unravel the three-dimensional (3D) electronic structure of CeRhIn$_5$;
ii) show that spin flip scattering in CeRhIn$_5$ results in the buildup  of Kondo resonances;
iii) demonstrate that resonant ARPES with polarized light allows for the extraction of orbital information of CEF levels from Kondo satellite peaks near $E_F$;
iv) find that the Ce $4f$-$c$-electron hybridization in CeRhIn$_5$ is band dependent with strongest hybridization on the most 3D band, and it is  the weakest compared with those of CeCoIn$_5$ and CeIrIn$_5$;
v) argue that the FS enlargement is minimal in the investigated $T$ range.

High quality single crystals of CeRhIn$_5$ were grown by the self-flux method~\cite{Jiao.15}.
The soft X-ray ARPES data in Figs.~1 and 2  were taken at the Advanced Resonant Spectroscopies (ADRESS) beamline at the Swiss Light Source; the overall energy resolution was 70-80~meV, and the angular resolution was 0.07$^{\circ}$. The samples were cleaved and measured at 12~K under a vacuum better than $5\times 10^{-11}$~mbar.
The resonant photoemission data  were taken with  121~eV photons  at Beamline I05-ARPES at the Diamond Light Source. The angular resolution was $0.2^{\circ}$ and the overall energy resolution was  better than 17 meV. The samples were cleaved at 8~K, and the vacuum was below $9\times10^{-11}$~mbar.
\begin{figure}[tbp]
\includegraphics[width=\columnwidth]{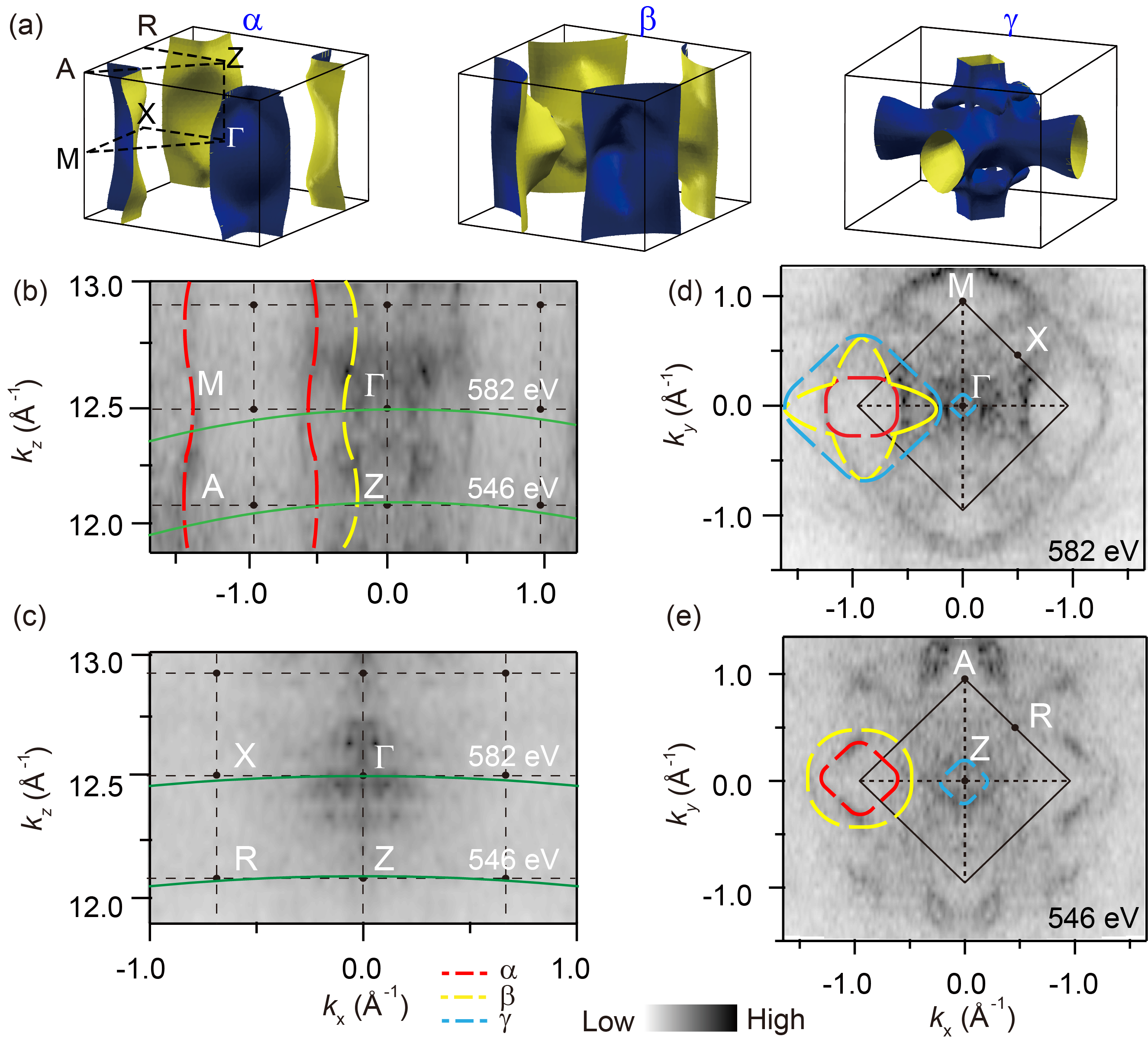}
\caption{FSs of CeRhIn$_5$ at 12 K. (a) Calculated  $\alpha$, $\beta$ and $\gamma$ FSs of CeRhIn$_5$, when considering fully localized $f$ electrons \cite{Jiao.15}. (b) and (c)  Photoemission intensity maps in the (b) $\Gamma ZAM$ and (c) $\Gamma ZRX$ plane. The dispersion of $\gamma$ is not marked due to its pronounced $k_z$ dependence.The momentum cuts taken with 546 and 582 eV has been marked with the green solid line in (b) and (c). (d) and (e) Photoemission intensity maps in the (d) $\Gamma XM$ and (e) $ZAR$ plane taken with 582 and 546~eV photons. All data  were integrated over a window of ($E_F$-20 meV, $E_F$+20 meV).
}
\label{FS}
\end{figure}

\begin{figure*}
\includegraphics[width=\textwidth]{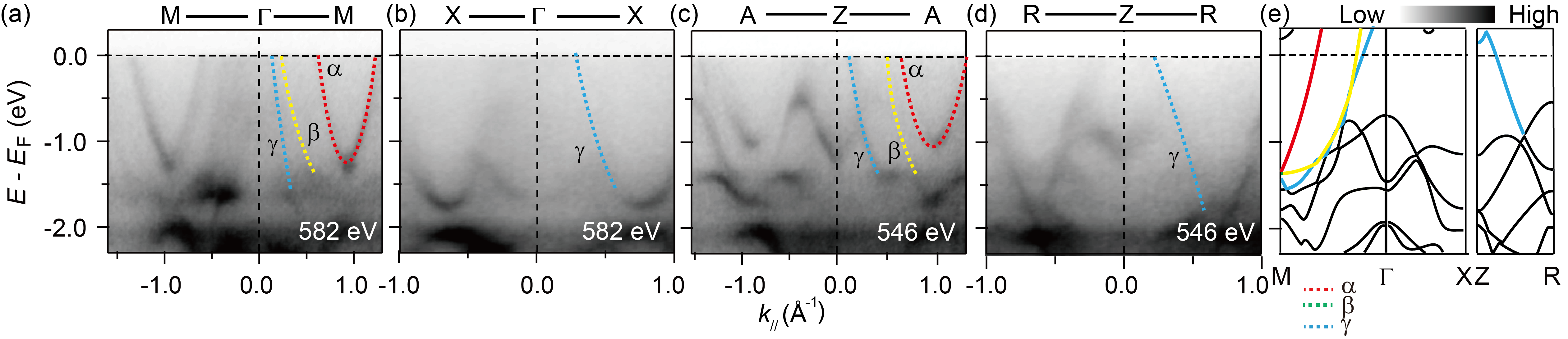}
\caption{Band structure of CeRhIn$_5$ at 12 K. (a-d) Photoemission intensity distributions along (a) $\Gamma$$M$, (b) $\Gamma$$X$, (c) $Z$$A$, and (d) $Z$$R$. The corresponding momentum cuts has been marked in Figs. 1(b) and (c). (e) Calculated band structure of CeRhIn$_5$ along high-symmetry directions, when considering fully localized $f$ electrons and a paramagnetic phase. The bands are color coded as marked. The positions of the momentum cuts are marked in Figs.~1(b) and 1(c).}
\label{band}
\end{figure*}



\begin{figure}[tbp]
\includegraphics[width=\columnwidth]{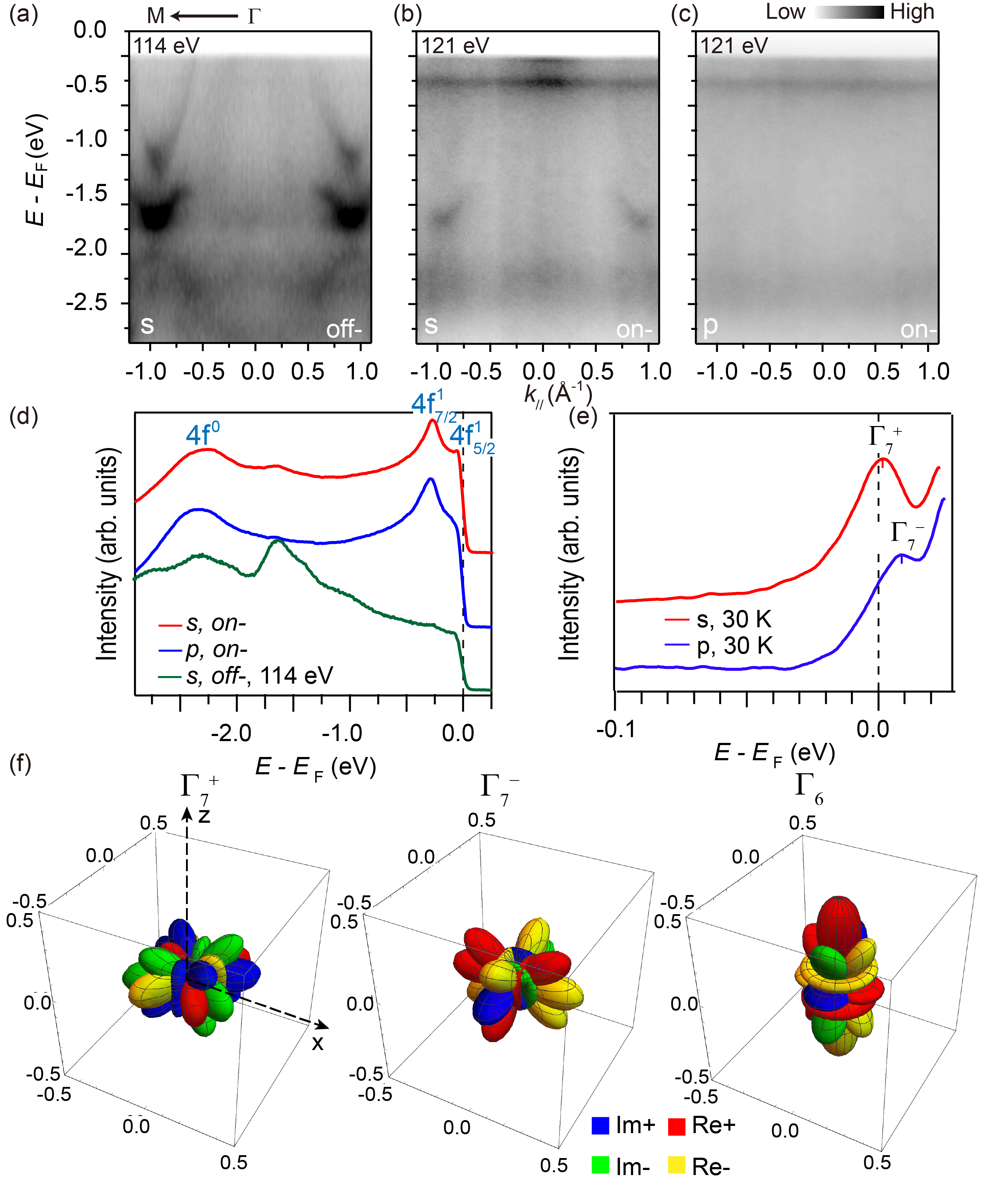}
\caption{Photoemission intensity distributions of CeRhIn$_5$  along $\Gamma$$M$ taken at 8 K with (a) off-resonance (114 eV)  $s$-polarized, (b) on-resonance (121~eV) $s$-polarized, and  (c)  on-resonance $p$-polarized photons. (d) Angle-integrated EDCs of CeRhIn$_5$ taken with off- and on-resonant energy; $f$-band positions are highlighted.
(e) On-resonance EDCs of CeRhIn$_5$ taken  at $\Gamma$  with  $s$- and $p$-polarized photons, after division by the RC-FDD.
Labels follow the common notation in Refs.~\cite{Im.08,Fujimori.03,Patil.16}. Momentum cuts with 121~eV photons cross (0,0,7.07$\frac{2\pi}{c}$), close to $\Gamma$, and are thus labeled $\Gamma$-M for simplicity.
(f)  The distributions of the wavefunctions for the three doublets in the tetragonal crystal electric field, named $\Gamma_7^{+}$, $\Gamma_7^{-}$, and $\Gamma_6$, respectively. The positive and negative real and imaginary parts of the wavefunctions are color coded as marked.}
\label{resonantPES}
\end{figure}


The 3D electronic structure of CeRhIn$_5$ was measured with soft x-ray ARPES, and compared with calculations when taking $f$ electrons as fully localized in Fig. 1(a).
Photoemission intensity maps in the $\Gamma ZAM$ and $\Gamma ZRX$ planes are shown in Figs.~1(b) and 1(c), respectively. Different $k_z$s were accessed by varying the photon energy between 532 and 648 eV, and estimated based on an inner potential of 15~eV \cite{SI}. As elaborated in the supplemental material \cite{SI}, according to the ``universal curve'' of electron mean free path as a function of kinetic energy \citep{Fujimori.16}, the probing depth here is larger than $\sim$14~\AA,  and the $k_z$-broadening is less than $\sim$8.5\%~$2\pi/c$, $c$ being the lattice constant of CeRhIn$_5$, which enables a proper measurement of the bulk 3D band structure \cite{SI,Xia.09}.
The FS in the $\Gamma XM$ plane, shown in Fig.~1(d), is composed of four pockets. There are three electron pockets around $M$: a flower-shaped $\beta$ pocket, a rounded $\alpha$ pocket and a large square-like $\gamma$ pocket. A small hole  pocket can be observed around $\Gamma$, which is attributed to the $\gamma$ band (Fig.~\ref{band}(b)). As shown by the band structure along $\Gamma$$M$ in Fig.~2(a), $\alpha$ appears V-shaped with its bottom 1.3~eV below $E_F$. In the $ZAR$ plane (Fig.~1(e)), the $\beta$ pocket becomes rounded and the $\alpha$ pocket becomes square-like. For the photoemission data along  $Z$$A$ in Fig.~2(c), the bottom of $\alpha$ shifts toward $E_F$, while $\beta$ gets closer to $\alpha$. As shown in Figs.~2(b) and 1(d), $\gamma$ encloses the $\Gamma$ point forming the square-like Fermi pocket around the zone center.

Differences exist between the FSs  in the $\Gamma XM$ and $ZAR$ planes for the three bands.
The $\alpha$ and $\beta$ FSs show weak variation in the $\Gamma ZAM$ plane (Fig.~1(b)), and $\alpha$ is the most two-dimensional (2D), while $\gamma$ exhibits the strongest $k_z$ dependence. The shapes of the $\alpha$ and $\beta$ pockets   are even qualitatively compatible with previous dHvA measurements in the magnetic phase
\cite{Shishido.02}.
These measurements, however,  could not provide the shape and $k_z$ dependence of $\gamma$, due to the low frequency of this branch, making the more comprehensive 3D electronic structure revealed by ARPES particularly useful.
Most parts of the FSs and bands agree well with the DFT calculations for the paramagnetic phase, shown in Fig.~1(a) and Fig.~2(e), where the 4$f$ electrons were treated fully localized and  Kondo spin scattering is neglected. This should reproduce the  FSs when the effect of the coupling between 4$f$ electrons and conduction electrons is negligible.
This agreement indicates that the 4$f$ electrons indeed are predominantly localized. Still, differences between the DFT calculations and our measurements exist: the Fermi crossing of $\gamma$ along $\Gamma X$ in Fig.~2(b) is absent in Fig.~2(e).
The calculation also indicates all  three conduction bands  originate from the RhIn$_2$ layer, with $\alpha$ mainly composed of Rh 4$d$ states, $\beta$ mainly composed of In 5$p$ states, and $\gamma$ composed of both.

To enhance the $f$-electron photoemission intensity, resonant ARPES measurements were conducted at the Ce 4$d$-4$f$ transition with 121 eV photons. As shown in Fig. S2 of the supplemental material \cite{SI}, the data taken with 882.5 eV photons at the 3$d$-4$f$ resonance are similar, but with a poor energy resolution. As further illustrated by  Figs.~S3 and S4 \cite{SI}, the  photons near 4$d$-4$f$ transition probe the bulk states as well, although the probing depth is about 7.5~\AA, causing stronger $k_z$-broadening.
Figs.~3(a) and 3(b) compare the off-resonance and on-resonance photoemission intensities at 8 K, respectively. Strongly dispersing bands dominate the off-resonance spectra, while Ce 4$f$ emission is enhanced in the on-resonance data, see also Fig.~3(d). Three nearly flat bands can be observed in the on-resonance data. The one at -2.32 eV  is assigned to the $4f^0$ state, while those at \mbox{-0.27 eV} and near $E_F$ have been attributed to the $4f_{7/2}^{1}$  and $4f_{5/2}^{1}$ states, respectively. The $4f$ spectral weight was found to be negligible near $E_F$  in previous ARPES data on CeRhIn$_5$  \cite{Fujimori.03}.
Furthermore, we find that $4f_{5/2}^{1}$ is sensitive to the polarization -- a significant enhancement is seen under $s$-polarized light compared with $p$ polarization (Fig.~3(c)).
Fig.~3(e) enlarges the  energy distribution curves (EDCs) near $E_F$ after dividing by the resolution-convoluted Fermi-Dirac distribution (RC-FDD), revealing two features at $\sim$1.5 and 8~meV which we interpret as Kondo satellite peaks arising from the CEF splittings.
The polarization dependence arises from wavefunction symmetry. The $4f_{5/2}^{1}$ manifold splits into three doubly-degenerate states in the tetragonal environment, namely
$\Gamma_7^{+}= \beta\left|\pm\frac{5}{2}\right> - \alpha\left|\mp\frac{3}{2} \right>$,
$\Gamma_6=   \left|\pm\frac{1}{2}\right>$, and
$\Gamma_7^{-}=\alpha\left|\pm\frac{5}{2} \right> +\beta\left|\mp\frac{3}{2}\right>$.
Here, $\alpha$=0.62 and $\beta$=0.78 were determined by inelastic neutron scattering \cite{Willers.10}. The corresponding distributions of these wavefunctions are displayed in Fig. 3(f). All these are somewhat elongated along $z$, allowing hybridization with the conduction electrons.
Both $\Gamma_7^{+}$ and $\Gamma_7^{-}$ have mixed even and odd components with respect to the $xz$ plane (defining the $x$-axis as the Ce--In--Ce direction and $z$ as the $c$-axis), while $\Gamma_6$ is even. Moreover, the odd component of $\Gamma_7^{+}$ is larger than its even component, while the reverse is true for $\Gamma_7^{-}$.
Since $s$-polarization picks up the odd components, the prominent state at 1.5~meV above $E_F$ is assigned to $\Gamma_7^{+}$. The Kondo satellite state observed with $p$-polarization at 8~meV above $E_F$  is assigned to $\Gamma_7^{-}$. The observed separation of 6.5 meV between the two states agrees well with that from inelastic neutron scattering \cite{Willers.10}.
The third state, $\Gamma_6$, is too far above $E_F$ (25 meV) to be resolved here.
Thus, ARPES with polarized light can be a valuable complement to neutron scattering and specific heat measurements in determining the CEF states in Kondo systems.
The $4f_{7/2}^{1}$ multiplet at higher binding energy features a broader lineshape than the $4f_{5/2}^{1}$ one. Individual Kondo satellite peaks therefore were not resolved.


\begin{figure}
\includegraphics[width=\columnwidth]{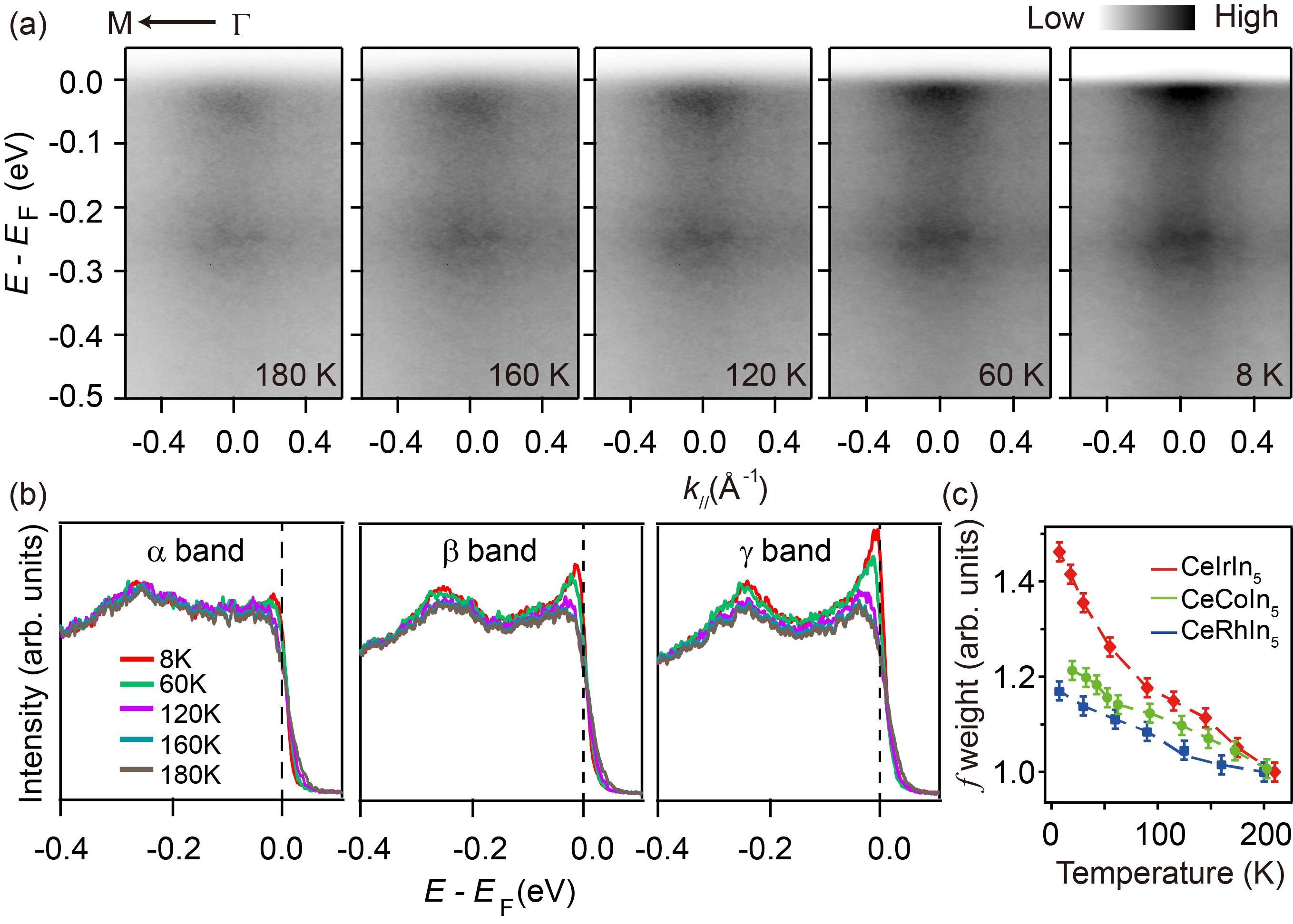}
\caption{Temperature evolution of the electronic structure of CeRhIn$_5$. (a) Resonant ARPES data along  $\Gamma$$M$ at the temperatures indicated. (b) $T$-dependence of the EDCs at the Fermi crossings of $\alpha$, $\beta$ and $\gamma$.  Thermal broadening near $E_F$ is removed by dividing the spectra with the RC-FDD at corresponding temperatures and then multiplied by that of 8 K. (c) $T$--dependence of the quasiparticle spectral weight near $E_F$ in the vicinity of $\Gamma$ for CeCoIn$_5$ \cite{Qiuyun.17}, CeRhIn$_5$, and CeIrIn$_5$ \cite{Qiuyun.unpl,SI}, integrated over [$E_F-$40~meV, $E_F+$20~meV].}
\label{tdep}
\end{figure}

To investigate the degree of $f$-electron delocalization, we performed $T$-dependent measurements along $\Gamma$$M$, see Fig.~4(a). At high $T$, only two weak structures are observed --- one at $E_F$, the other at around -0.27~eV. Upon cooling, the intensity of both features gradually increases until significant spectral weight can be observed near $E_F$ at low $T$, particularly near  $\Gamma$.
This further corroborates the interpretation of both features as Kondo resonances which in principle should be detectable using STS~\cite{Aynajian.12,SI}.

Figure 4(b) displays the EDCs at the Fermi crossings of $\alpha$, $\beta$ and $\gamma$ at low temperature. The 4$f$-$c$-electron hybridization
displays a clear band-dependence --
 the most 2D $\alpha$ band shows negligible hybridization, while  the most 3D $\gamma$ band shows the strongest hybridization. This is consistent with both thermal expansion and DMFT results \cite{Willers.10,Oeschler.03}.
Consistently, a nearly flat band displaying a weak dispersion near $\Gamma$ is clearly visible in the EDCs taken at 8 K near $E_F$ due to the hybridization between the 4$f$ band and  $\gamma$, see Fig. 5(a).
Fig. 4(c) compares the $T$-dependence of $f$ spectral weight in the vicinity of $\Gamma$ of CeRhIn$_5$ and those of its sister compounds CeCoIn$_5$ and CeIrIn$_5$. Obviously, the $f$ spectral weight in CeRhIn$_5$ increases  slower upon lowering $T$ than those in CeCoIn$_5$ and CeIrIn$_5$, indicating that the hybridization is weakest in CeRhIn$_5$ among these compounds.
In the supplemental material \cite{SI}, we compare the band structure of the three compounds. An obvious band bending can be observed in $\alpha$  for CeCoIn$_5$ and CeIrIn$_5$, but not  for CeRhIn$_5$ down to 8 K. In short, the 4$f$-$c$ hybridization  weakens in the order of CeIrIn$_5$,  CeCoIn$_5$, and CeRhIn$_5$.

Having firmly established that Kondo resonances do form in CeRhIn$_5$ as $T$ is lowered, we now turn to a discussion of the size of the FS.
It is important to note that the formation of Kondo resonances is not \textit{per se} tantamount to the formation of a large FS.
As sketched in Fig.~5(b),  {\it i.e.}, according to the standard view on heavy fermion groundstate,  when $f$ electrons hybridize with conduction electrons, the intensity of the $f$ spectral weight will be redistributed, with significant enhancement to the `inside' of the hole-like and `outside' of the electron-like bands. This is  demonstrated by comparing the  photoemission intensity maps of CeRhIn$_5$  taken at 8 and 180~K, respectively, in Figs.~5(c) and 5(d).
The change of the Fermi wavevector from $k_F$,  the Fermi wavevector of the so-called `small' FS with the $f$-moments being localized, to $k_F'$ of the `large' FS  entails a reduction of spectral weight at  $k_F$ according to Fig.~5(b). Although there are signs for hybridization in Fig.~4 and Fig.~5(a),
this reduction is not observed down to the lowest $T$ of our experiment, see Fig.~5(c). Therefore,
our findings indicate that the enlargement of the Fermi surface is minimal, and it is still predominantly
`small' in CeRhIn$_5$ at ambient pressure down to 8~K  despite the formation of a flat band near $E_F$ ~\cite{Shishido2005,Jiao.15},
and also clarify why some experiments on CeRhIn$_5$ have been interpreted in terms of complete $4f$-electron delocalization and do show how subtle the issue of large vs small FS can be.

\begin{figure}
\includegraphics[width=\columnwidth]{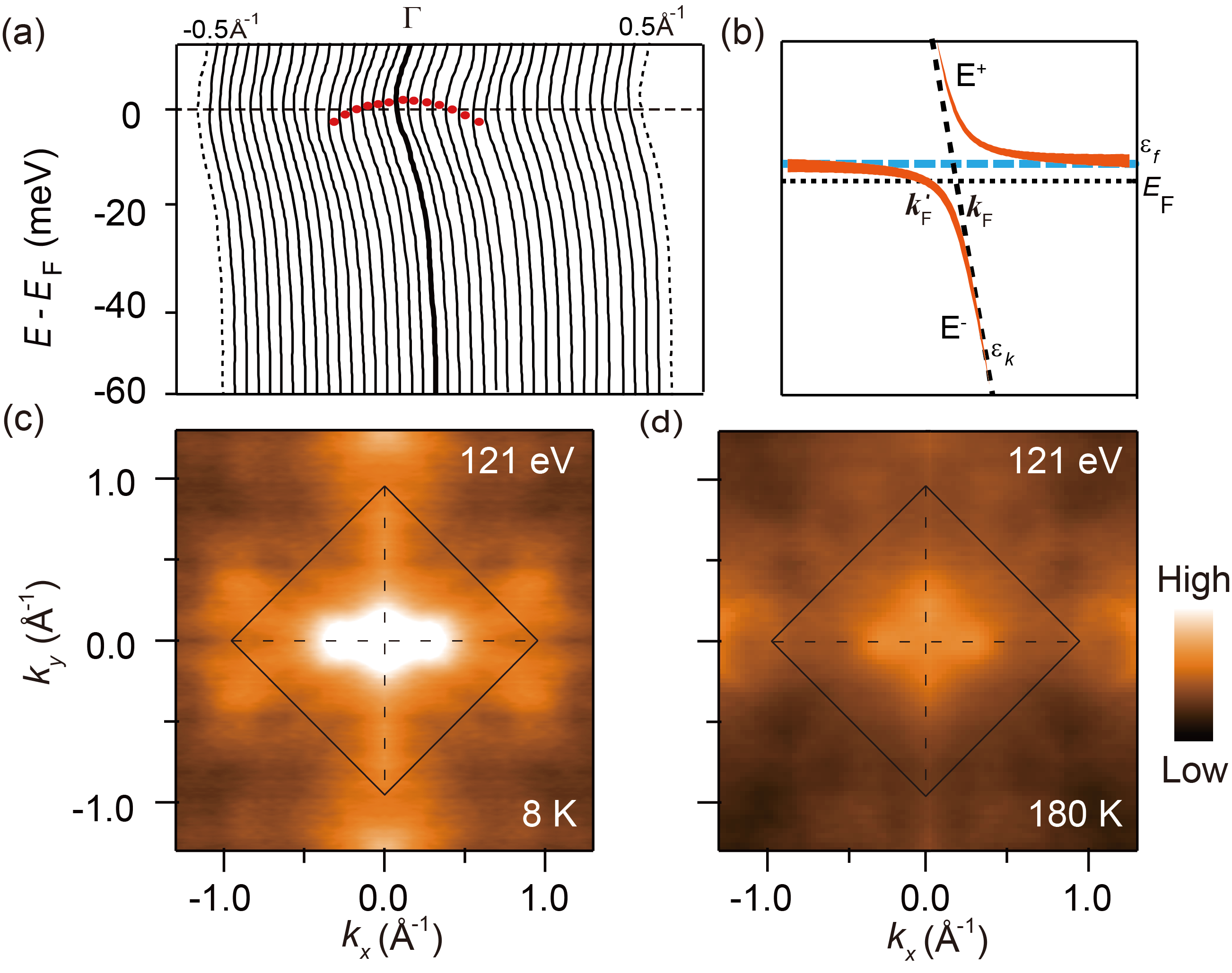}
\caption{Formation of the heavy quasiparticle band in CeRhIn$_5$. (a) The EDCs  at 8 K along $\Gamma$$M$, divided by the RC-FDD. (b) Schematic diagram of the hybridization between $f$ electrons  ($\varepsilon_f$) and a conduction band ($\varepsilon_k$) under a periodic Anderson model. The orange curve is the hybridized band. (c,d) Photoemission intensity maps at (c) 8 and (d) 180~K.}
\label{fig5}
\end{figure}

In conclusion, we demonstrate that  polarization-dependent ARPES could extract information about the CEF levels in CeRhIn$_5$. We showed that there is band-dependent 4$f$-c hybridization, and
the FS in the paramagnetic phase of CeRhIn$_5$ is predominantly small down to 8 K, although a narrow band forms near $E_F$ as a result of Kondo scattering. These findings clarify why different experiments have seemingly obtained conflicting results regarding the fate of the Kondo effect in CeRhIn$_5$.
Our findings give a comprehensive  electronic structure of CeRhIn$_5$,  and the comparison with CeCoIn$_5$  and CeIrIn$_5$  aides a global understanding of the Ce$M$In$_5$ family
which epitomizes the richness of strongly correlated electron systems~\cite{note3}.

\begin{acknowledgments}
We gratefully acknowledge enlightening discussions with Prof.\ P.\ Coleman and Prof.\ Y.\ F.\ Yang, and the experimental support of Dr.\ M.\ Hoesch and Dr.\ T.\ Kim. This work is supported in part by the National Key Research and Development Program of China (Grant No.\ 2016YFA0300200 and No. 2017YFA0303104), the National Science Foundation of China (Grants No.\ 11504342, 11474250, and No. U1630248), Science Challenge Project (No.\ TZ2016004),  the Science and Technology Commission of Shanghai Municipality (Grant No.\ 15ZR1402900), Diamond Light Source for time on beamline I05 under Proposal No.\ SI11914. Soft x-ray ARPES was performed at the SX-ARPES endstation of the ADRESS beamline at the Swiss Light Source, Paul Scherrer Institute, Switzerland. Some preliminary data were taken at the ARPES beamline of Shanghai Synchrotron Radiation Facility (SSRF, China) and the National Synchrotron Radiation Laboratory (NSRL).
\end{acknowledgments}


\begin{thebibliography}{10}
\bibitem{note3}It will be interesting to address how our findings relate to the intriguing finite field behavior of CeRhIn$_5$~\cite{Ronning.17}.
\end{thebibliography}
\end{document}